\documentclass[prb,twocolumn,showpacs,amsmath,amssymb]{revtex4}

\usepackage{graphicx}% Include figure files
\usepackage{dcolumn}% Align table columns on decimal point
\usepackage{bm}% bold math
\usepackage{ifthen}

\begin{document}

\title{Excitonic exchange effects on the radiative decay time of monoexcitons
  and biexcitons in quantum dots}

\author{Gustavo A. Narvaez}
\altaffiliation{Current address: Eclipse Energy Systems, Inc., St. Petersburg, Florida 33710}
\email{gnarvaez@eclipsethinfilms.com}
\author{Gabriel Bester}
\author{Alberto Franceschetti}
\author{Alex Zunger}
\email{alex_zunger@nrel.gov}
\affiliation{National Renewable Energy Laboratory, Golden, Colorado 80401}
\date{\today{}}
\begin{abstract}
  Electron-hole exchange interactions split the exciton ground state into
  ``dark" and ``bright" states. The dynamics of those states depends on the
  internal relaxation time between bright and dark states (spin-flip time),
  and on the radiative recombination time of the bright states.
  On the other hand, the calculated values of these recombination times depend
  not only on the treatment of correlation effects, but also on the accuracy
  of the electron and hole wavefunctions.
We calculate the radiative decay rates for monoexcitons and biexcitons in
(In,As)Ga/GaAs self-assembled and colloidal CdSe quantum dots from atomistic
correlated wave functions. 
  We show how the radiative decay time $\tau_R(X^0)$ of the monoexciton
  depends on the spin-flip relaxation time between bright and dark states. In
  contrast, a biexciton has no bright-dark splitting, so the decay time of the
  biexciton $\tau_R(XX^0)$ is insensitive to this spin-flip time.
  This results in ratios $\tau_R(X^0)/\tau_R(XX^0)$ of 4 in the
  case of fast spin flip, and a ratio of 2 in the case of slow spin flip.
  For (In,Ga)As/GaAs, we compare our results with the model calculation of
  Wimmer {\em et al.} [M. Wimmer {\em et al.}, Phys. Rev. B {\bf 73}, 165305
  (2006)]. When the same spin-flip rates are assumed, our predicted
  $\tau_R(X^0)/\tau_R(XX^0)$ agrees with that of Wimmer {\em et al.},
  suggesting that our treatment of correlations is adequate to predict the
  ratio of monoexciton and biexciton radiative lifetimes.
  Our results agree well with experiment on self-assembled quantum dots when
  assuming slow spin flip. Conversely, for colloidal dots the agreement with
  experiment is best for fast spin flip.
\end{abstract}


\maketitle

\section{Introduction: Relation between apparent and microscopic carrier decay}

\begin{figure}
\includegraphics[width=7.5cm]{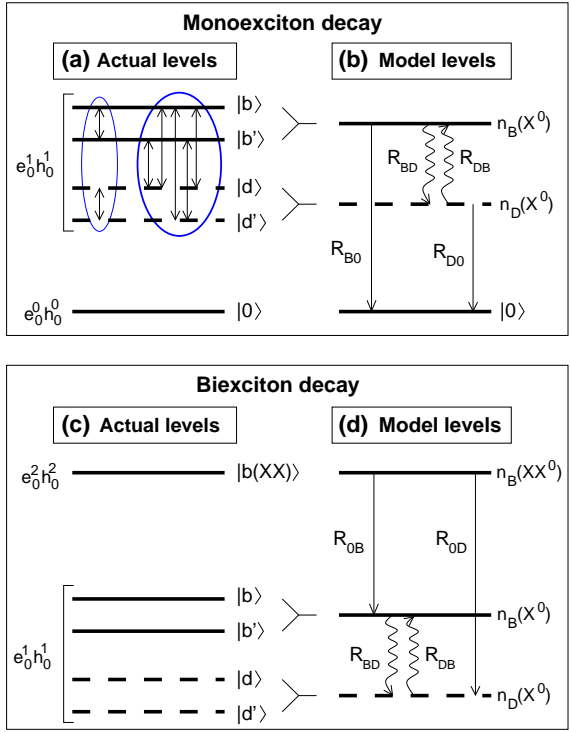}
\caption{{\label{Fig_1}}(Color online.) Sketch of (a) the four states that 
  encompass the monoexciton ground state $e^1_0h^1_0$ and $e_0^0h_0^0$. Thin-
  and thick-line  ellipses indicate, respectively, non-radiative thermalization as well as spin-flip channels.
  (b) The model three-level system for the monoexciton decay. $R_{BD}$
  ($R_{DB}$) is the bright-dark (dark-bright) rate while $R_{B0}$ and $R_{D0}$
  are radiative decay rates for the bright and dark (model) state,
  respectively. (c) {\em Idem (a)} for the biexciton ground state $e_0^2h_0^2$
  and $e^1_0h_0^1$, and (d) model representation, with radiative decay rates
  $R_{0B}$ and $R_{0D}$.}
\end{figure}

We address here the subject of how to compare measured exciton $\tau_{R}(X^0)$
and biexciton $\tau_{R}(XX^0)$ radiative relaxation times with calculated
values.
Experimentally, an ensemble of quantum dots is excited by an optical
pump-pulse and the photons subsequently emitted are counted as a function of
time.
The photon emission rate {\em vs} time is often not a simple exponential. The
reason for this is that even in a single dot the monoexciton ground state is
not a single state but a manifold of exchange and fine-structure split states
with internal carrier dynamics. In III-V and II-VI dots the monoexciton ground
state originates from $e^1_0h^1_0$, where $e_0$ and $h_0$ are, respectively,
the lowest- and highest-energy confined electron and hole states.
Due to the electron-hole exchange interaction, this state is not four fold
degenerate but splits into four lines [Fig. \ref{Fig_1}(a)].
\cite{bayer_FSS,bester_PRB_2003,franceschetti_PRB_1999}
For the $C_{2v}$ symmetry of (In,Ga)As/GaAs self-assembled quantum dots the
four states are the high-energy bright state $B$ consists of a pair $b$ and
$b^{\prime}$ split by a few $\mu{\rm eV}$ while the low-energy dark state $D$
consists of a pair $d$ and $d^{\prime}$ that is quasi-degenerate. $B$ and $D$
are split by a few hundred $\mu{\rm eV}$ due to exchange effects.
For colloidal CdSe quantum dots with the $C_{6v}$ symmetry the internal
$d$-$d^{\prime}$ and $b$-$b^{\prime}$ splittings of the two pairs is small,
with the $b$-$b^{\prime}$ recently measured to be about $1$-$2\;{\rm meV}$.\cite{furis_condmat_2005} In turn, the $b$-$d$ splitting between the
dark and the bright states is an order of magnitude larger than in
self-assembled dots, ranging from $2$-$20\;{\rm meV}$.\cite{nirmal_PRL_1995,furis_condmat_2005}
% Biexciton
%
In both self-assembled and colloidal dots, the biexciton ground state
$e^2_0h^2_0$ state has no fine structure and corresponds to a single bright
state that can decay to the four states of $e^1_0h^1_0$ in the monoexciton
[Fig. \ref{Fig_1}(c)].

In this paper we show that (i) due to exchange and fine-structure in the
monoexciton, the measured apparent radiative recombination time
$\tau_{R}(X^0)$ depends on the bright-to-dark spin-flip relaxation time
$\tau_{BD}$ with rate $R_{BD}=\tau^{-1}_{BD}$. 
By using an atomistic pseudopotential-based approach combined with the
configuration-interaction method,\cite{zunger_pssb_2001} we calculate the
characteristic radiative recombination rates between $B$ and the ground state
($R_{B0}$) and between $D$ and the ground state ($R_{D0}$) and input them in a
set of rate equations with varying $R_{BD}$ rates. We find that the photon
emission rate decays as a single-exponential with rate $\simeq R_{B0}$ for
slow spin flip times; as a biexponential for intermediate $\tau_{BD}$; and as
a single-exponential with rate $R_{B0}/2$ for fast spin flip times.
(ii) Within the same approach used for the monoexciton, we calculate the
characteristic recombination rates $R_{0B}$ and $R_{0D}$ of the biexciton
ground state into the bright and dark states of the monoexciton. We find that
$R_{0B}\simeq R_{B0}$ and $R_{0D}\simeq R_{D0}$, and that the biexciton
radiative decay is a single exponential with a decay time $\tau(XX^0)\simeq
2R^{-1}_{0B}$ regardless of $R_{BD}$.
(iii) We show that due to the aforementioned dependence of the monoexciton
decay time on $R_{BD}$, the ratio $\tau_{R}(X^0)/\tau_R(XX^0)$ has the values
of 4 and 2 for the limiting cases of fast and slow spin flip, respectively.
We thus resolve the apparent contradiction between the recent model
calculations of Wimmer and co-workers\cite{wimmer_PRB_2006}, who found
$\tau_{R}(X^0)/\tau_R(XX^0) \simeq 2$, and our previous atomistic-based
realistic calculations of $\tau_{R}(X^0)$ and $\tau_R(XX^0)$ in which we
found\cite{narvaez_PRB_2005c} $\tau_{R}(X^0)/\tau_R(XX^0) \simeq 4$.
We illustrate our findings with atomistic, pseudopotential-based calculations
for a prototypical self-assembled In$_{\rm 0.6}$Ga$_{\rm 0.4}$As/GaAs dot and
a CdSe colloidal dot, comparing with available data.

%%%%%%%%%%%%%%%%
%   Figure 1   %
%%%%%%%%%%%%%%%%
%
%

%%%%%%%%%%%%%%%%%%%%%%
%                    %
%   Rate equations   %
%                    %
%%%%%%%%%%%%%%%%%%%%%%
%
%
\section{Rate equations for the radiative decay of the monoexciton}

Figure \ref{Fig_1}(a) shows the monoexciton and biexciton enery levels that
enter our calculations. We do not consider higher-lying states because most
time-resolved photoluminescence experiments are conducted at temperatures such
that the occupation of those states is negligible.
From Fig. \ref{Fig_1}(a) we see that there are a number of discrete
transitions channels denoted below by rates $R_{ij}$. We next set up a set of
channel-specific rate equations describing how the individual levels of Fig.
\ref{Fig_1}(a) ``communicate,'' from which we will deduce the global decay of
the ground state $n_0(t)$ which is measured.
Using the characteristic radiative rates of the four excitonic states of
$e^1_0h^1_0$ and the ground state $e^0_0h^0_0$ we establish the following
system of rate equations:

\begin{widetext}
\begin{eqnarray}
\label{eq:5levels}
{ d}n_b/{ d}t &=& -\left( R_{bb^{\prime}} + R_{bd} + R_{bd^{\prime}} + R_{b0}\right) n_b + R_{d^{\prime}b}\,n_{d^{\prime}} + R_{db}\,n_d + R_{b^{\prime}b}\,n_{b^{\prime}}  \nonumber \\ 
{ d}n_{b^{\prime}}/{ d}t &=& -\left( R_{b^{\prime}b} + R_{b^{\prime}d} + R_{b^{\prime}d^{\prime}} + R_{b^{\prime}0}\right) n_{b^{\prime}} + R_{d^{\prime}b^{\prime}}\,n_{d^{\prime}} + R_{db^{\prime}}\,n_d + R_{bb^{\prime}}\,n_b  \nonumber \\ 
{ d}n_d/{ d}t &=& -\left( R_{db} + R_{d^{\prime}b^{\prime}} + R_{dd^{\prime}} + R_{d0}\right) n_d + R_{d^{\prime}d}\,n_{d^{\prime}} + R_{b^{\prime}d}\,n_{b^{\prime}} + R_{bd}\,n_b  \\
{ d}n_{d^{\prime}}/{ d}t &=& -\left( R_{14} + R_{13} + R_{12} + R_{10}\right) n_{d^{\prime}} + R_{dd^{\prime}}\,n_{d} + R_{b^{\prime}d^{\prime}}\,n_{b^{\prime}} + R_{bd^{\prime}}\,n_b \nonumber \\
{ d}n_0/{ d}t &=&  R_{d^{\prime}0}\,n_{d^{\prime}} + R_{d0}\,n_d + R_{b^{\prime}0}\,
n_{b^{\prime}} + R_{b0}\,n_b , \nonumber
\end{eqnarray}
\end{widetext}

%%%%%%%%%%%%%%
%   Table 1  %
%%%%%%%%%%%%%%
%
%
\begin{table*}[t]
\caption{{\label{Table_1}}Calculated values for the radiative characteristic
    rates $R_{i0}$ and $R_{bi}$ $(i=b,\,b^{\prime},\,d,\,d^{\prime})$ for the
    monoexciton ($X^0$) and biexciton ($XX^0$), respectively, in an alloyed In$_{\rm
    0.6}$Ga$_{\rm 0.4}$As/GaAs dot (base diameter $b=252\;${\AA} and
    height $h=35\;${\AA}) and a CdSe colloidal dot (diameter
    $D=38\;${\AA}). The approximate rates $R_{B0}$, $R_{D0}$ that enter the
    model 3-level system of rate equations [Eq. (\ref{Eq_3})] are also shown,
    as well as the rates $R_{0B}$ and $R_{0D}$ for the biexciton.}

\begin{tabular}{ccccc}
\hline\hline
 & \multicolumn{2}{c}{In$_{\rm 0.6}$Ga$_{\rm 0.4}$As/GaAs} & \multicolumn{2}{c}{CdSe} \\ \hline 

$X^0$ & 
\begin{tabular}{ll}
$R_{b0}$ & $0.89\;{\rm ns}^{-1}$ \\
$R_{b^{\prime}0}$ & $0.91\;{\rm ns}^{-1}$ \\
$R_{d0}$ & $0.65\;10^{-5}\;{\rm ns}^{-1}$ \\
$R_{d^{\prime}0}$ & $0.53\;10^{-4}\;{\rm ns}^{-1}$
\end{tabular} & \begin{tabular}{l} 
$R_{B0}=0.9\;{\rm ns^{-1}}$ \\
\\
$R_{D0}=0$
\end{tabular} & \begin{tabular}{ll} $R_{b0}$ & $0.12\;{\rm ns^{-1}}$ \\
  $R_{b^{\prime}0}$  & $0.12\;{\rm ns^{-1}}$ \\
 $R_{d0}$  & $0.04\;10^{-6}\;{\rm ns^{-1}}$ \\ $R_{d^{\prime}0}$ &
 $0.22\;10^{-6}\;{\rm ns^{-1}}$ \end{tabular} 
& 
\begin{tabular}{l}
 $R_{B0}=0.12\;{\rm ns^{-1}}$ \\ \\ $R_{D0}=0$ 
\end{tabular} 
\\ \hline
$XX^0$ &
\begin{tabular}{ll}
$R_{bb}$ & $0.83\;{\rm ns}^{-1}$ \\
$R_{bb^{\prime}}$ & $0.85\;{\rm ns}^{-1}$ \\
$R_{bd}$ & $0.67\;10^{-5}\;{\rm ns}^{-1}$ \\
$R_{bd^{\prime}}$ & $0.46\;10^{-4}\;{\rm ns}^{-1}$ 
\end{tabular} & \begin{tabular}{l}
$R_{0B}=0.84\;{\rm ns^{-1}}$ \\ 
\\
$R_{0D}=0$
\end{tabular} & \begin{tabular}{ll} $R_{bb}$ & $0.13\;{\rm ns^{-1}}$ \\
  $R_{bb^{\prime}}$  & $0.13\;{\rm ns^{-1}}$ \\
 $R_{bd}$  & $0.09\;10^{-6}\;{\rm ns^{-1}}$  \\ 
$R_{bd^{\prime}}$ & $0.53\;10^{-6}\;{\rm ns^{-1}}$\end{tabular} 
& 
\begin{tabular}{l}
 $R_{0B}=0.13\;{\rm ns^{-1}}$ \\ \\ $R_{0D}=0$ 
\end{tabular} \\ 
\hline\hline
\end{tabular}
\end{table*}

\noindent where $R_{ij}$ are the characteristic recombination rates from the level
$i$ to level $j$. The five-level system of rate equations [Eq.
(\ref{eq:5levels})] that describe the radiative decay of $X^0$ can be reduced
to a three-level system when (i) the thermalization rate [Fig. \ref{Fig_1}(a)]
within the $b$-$b^{\prime}$ bright and within the $d$-$d^{\prime}$ dark states
is assumed equal:

\begin{equation}
\label{assumption_1}
R_{bb^{\prime}}=R_{b^{\prime}b}=R_{dd^{\prime}}=R_{d^{\prime}d}=R_{th};
\end{equation}

\noindent (ii) spin-flip rates between the dark and the 
bright states [Fig. \ref{Fig_1}(a)] are assumed to be independent on the index
of the bright or dark state while keeping the distinction between bright-dark
and dark-bright transition rates:

\begin{eqnarray}
\label{assumption_2}
& R_{bd}=R_{bd^{\prime}}=R_{b^{\prime}d}=R_{b^{\prime}d^{\prime}} = R_{BD} ,
\\
\label{assumption_3}
& R_{db}=R_{db^{\prime}}=R_{d^{\prime}b}=R_{d^{\prime}b^{\prime}} =  R_{DB}
;
\end{eqnarray}

\noindent and (iii) the decay rate of $b$ and $b^{\prime}$ to $e^0_0h^0_0=|0\rangle$
are equal, and so are the decays of $d$ and $d^{\prime}$ to $|0\rangle$:

\begin{equation}
\label{assumption_4}
R_{b0} = R_{b^{\prime}0} = R_{B0}, \quad R_{d0} = R_{d^{\prime}0} = R_{D0} .
\end{equation}

%
% Comment on assumptions for both SK and colloidal
%
\noindent Assumption (i) is justified in different range of temperatures 
determined by the magnitude of the small splittings between $b$-$b^{\prime}$
and $d$-$d^{\prime}$: $T\gtrsim 2\;{\rm K}$ in self-assembled dots and
$T\gtrsim 20\;{\rm K}$ in CdSe colloidal dots. 
Regarding (ii), the small fine-structure bright-dark splittings
($10$-$300\;\mu{\rm eV}$) in (In,Ga)As/GaAs dots allows us to make the
assumption $R_{BD}=R_{DB}$. For CdSe colloidal dots the bright-dark splitting
is about an order of magnitude larger
\cite{nirmal_PRL_1995,furis_condmat_2005} and therefore $R_{BD}=R_{DB}$ for
$T\gtrsim 30\;{\rm K}$. (In the results discussed in Sec.
\ref{integration_rate_eq} we adopt the regime in which $R_{BD}=R_{DB}$.)
Assumption (iii) is supported by our atomistic pseudopotential-based
calculations (see Sec. \ref{characteristic_times}). 
Assuming Eqs. (\ref{assumption_1})-(\ref{assumption_4}), we
simplify Eq. (\ref{eq:5levels}) to

\begin{eqnarray}
\label{Eq_3}
\frac{dn_B}{dt}&=& -(R_{B0}+2R_{BD})\,n_B+2R_{DB}\,n_D \nonumber\\
\frac{dn_D}{dt}&=& -(R_{D0}+2R_{DB})\,n_D+2R_{BD}\,n_B \\
\frac{dn_0}{dt}&=& R_{B0}\,n_B + R_{D0}\,n_D \nonumber ,
\end{eqnarray}

\noindent where $n_B=n_b+n_{b^{\prime}}$ is the occupation of the bright states and
$n_D=n_d+n_{d^{\prime}}$ the occupation of the dark states.  
We will calculate $R_{B0}$ and $R_{D0}$ from the electronic structure of the
dot and vary $R_{BD}$ in a wide range from very fast to very slow spin-flip
rates to examine the different regimes of behavior.
We will then solve Eq. (\ref{Eq_3}) and calculate the phonon emission rate
$I(t)=R_{B0}\,n_B(t)+R_{D0}\,n_D(t)$, which is directly comparable to
time-resolved photoluminescence (PL) experiments.

%%%%%%%%%%%%%%%%%
%   Biexciton   %
%%%%%%%%%%%%%%%%%
%
%
%
\section{Rate equation for the radiative decay of the biexciton}

The biexciton has a non-degenerate state without $B$-$D$ splitting and it
decays into the bright and dark states of the monoexciton. This decay can be
modeled, similarly to the monoexciton decay, with a three-level system [Fig.
\ref{Fig_1}(d)], yielding a single rate equation describing the population of
the biexciton ground state:

\begin{equation}
\label{rate_biexciton}
\frac{dn_B(XX^0)}{dt}=-2(R_{0B}+R_{0D})\,n_B(XX^0).
\end{equation}

\noindent As in the monoexciton case, we will calculate $R_{0B}$ and $R_{0D}$ 
from the electronic structure of the dot.

%%%%%%%%%%%%%%%%%%%%%%%%%%%%%%%%%%
%                                %
% Calculation of R_B0 and R_D0   %
%                                %
%%%%%%%%%%%%%%%%%%%%%%%%%%%%%%%%%%
%
%
%
\section{Calculation of bright and dark recombination rates from electronic structure}
\label{characteristic_times}

We use the empirical pseudopotential method where a superposition of screened
atomic pseudopotentials are used to describe the quantum dot
potential.\cite{zunger_pssb_2001} We take spin-orbit interaction into account
and the method naturally includes inter-band coupling and inter-valley
coupling.  Following the diagonalization of the single-particle Hamiltonian,
we use a configuration-interaction (CI) approach\cite{franceschetti_PRB_1999}
to obtain correlated monoexciton and biexciton wave functions
$|\Psi^{(\nu)}(\chi)\rangle$ ($\chi=X^0,XX^0$).
The characteristic radiative recombination rates $R_{if}(\chi)$ are calculated
using Fermi's golden rule from the correlated exciton wave functions as
follows. For a transition
$|\Psi^{(i)}(\chi)\rangle\rightarrow|\Psi^{(f)}(\chi-1)\rangle$,
$R_{if}(\chi)$ follows from both the magnitude of the dipole matrix element of
the transition $\big|{\bf M}^{(\hat{\bf e})}_{if}(\chi)\big|^2$ and the
recombination energy $\omega_{if}$. Namely,

\begin{equation}
\label{intrinsic.lifetime}
R_{if}(\chi^q)=\frac{4\,G}{3}\,\left(\frac{e^2}{m_0^2\,c^3\,\hbar^2}\right)n\,\omega_{if}(\chi^q)\sum_{\hat{{\bf
      e}}=\hat{x},\hat{y},\hat{z}}\big|{\bf M}^{(\hat{\bf e})}_{if}(\chi)\big|^2.
\end{equation}

\noindent Here, $e$ and $m_0$ are the charge and mass of the electron, 
respectively, and $c$ is the velocity of light in vacuum; the refractive index
$n$ of the dot material accounts for the material's effects on the photon
emission; and $G=G(\epsilon_{in},\epsilon_{out})$ accounts for
the dielectric constant mismatch between the dot material ($\epsilon_{in}$)
and medium ($\epsilon_{out}$)---solid barrier in self-assembled dots and
liquid solvent in colloidal.

Table \ref{Table_1} shows the calculated characteristic radiative
recombination rates $R_{i0}$ and $R_{bi}$ ($i=b,b^{\prime},d,d^{\prime}$) for
the monoexciton and biexciton, respectively, in a prototypical lens-shaped
In$_{\rm 0.6}$Ga$_{\rm 0.4}$As/GaAs self-assembled quantum dot with base
diameter $b=252\;${\AA} and height $h=35\;${\AA}, and a colloidal CdSe quantum
dot with diameter $D=38\;${\AA}.
We find that the rates for bright states indeed satisfy
$R_{b0}\simeq R_{b^{\prime}0}$ and that dark states obey $R_{d0}\sim
R_{d^{\prime}0} \sim 0$.

%%%%%%%%%%%%%%%%
%   Figure 2   %
%%%%%%%%%%%%%%%%
%
%
\begin{figure}
\includegraphics[width=8.0cm]{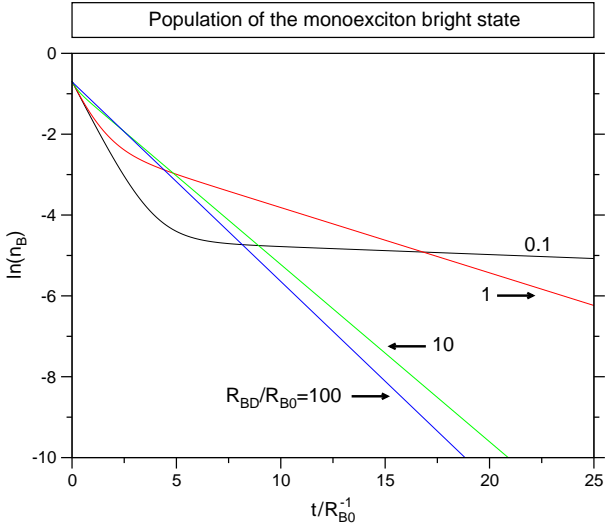}
\caption{{\label{Fig_2}} (Color online.) Population $n_B$
  [Eq. (\ref{general_form}); 
logarithmic scale] of the model bright state of the
  monoexciton versus time ($t$; units of $R^{-1}_{B0}$) for different 
  of the spin-flip rate $R_{BD}$. $R_{B0}=0.9{\rm ns^{-1}}$ for an
  In$_{\rm 0.6}$Ga$_{\rm 0.4}$As/GaAs quantum dot with base diameter
  $b=252\;${\AA} and height $h=35\;${\AA} and $R_{B0}=0.12\;{\rm ns^{-1}}$ for
  a CdSe dot with diameter $D=38\;${\AA} (Table \ref{Table_1}).}
\end{figure}

%%%%%%%%%%%%%%%%%%%%%%%%%%%%%%%%%%
%                                %
%   Integrating rate equations   %
%                                %
%%%%%%%%%%%%%%%%%%%%%%%%%%%%%%%%%%
%
%
%
\section{Analytic solution to the model rate equations for the monoexciton}
\label{integration_rate_eq}

The calculated radiative rates for the In$_{\rm 0.6}$Ga$_{\rm 0.4}$As/GaAs
self-assembled dot and the CdSe colloidal dot (Table \ref{Table_1}) show that
it is a good approximation to consider $R_{D0}=0$.
In this case, together with the assumptions that $R_{BD}=R_{DB}$
and\cite{dalgarno_pssa_2005} $n_B(0)=n_D(0)=1/2$, the solution of the model
three-level system of rate equations [Eq.  (\ref{Eq_3})] gives the following
(the general framework for the analytic results is in Appendix
\ref{analytic_solution}):

\begin{equation}
\label{general_form}
n_B(t) = F\exp(-\gamma_F\,t)+S\exp(-\gamma_S\,t) 
\end{equation}

\noindent with

\begin{eqnarray}
\label{gamma_A.solution.Table_1}
\gamma_{F}&=&\frac{1}{2}(R_{B0}+4R_{BD})+\frac{1}{2}\sqrt{R_{B0}^{\,2}+(4R_{BD})^2}
, \\
\label{gamma_B.solution.Table_1}
\gamma_{S}&=&\frac{1}{2}(R_{B0}+4R_{DB})-\frac{1}{2}\sqrt{R_{B0}^{\,2}+(4R_{BD})^2}
\end{eqnarray}

\noindent and

\begin{eqnarray}
\label{amp_A.solution.Table_1}
F& = &\frac{1}{2}\left(\frac{R_{B0}-\gamma_S}{\gamma_F-\gamma_S}\right) , \\
\label{amp_B.solution.Table_1}
S & = & -\frac{1}{2}\left(\frac{R_{B0}-\gamma_F}{\gamma_F-\gamma_S}\right) .
\end{eqnarray}

%%%%%%%%%%%%%%%%
%   Figure 3   %
%%%%%%%%%%%%%%%%
%
%
\begin{figure}
\includegraphics[width=8.5cm]{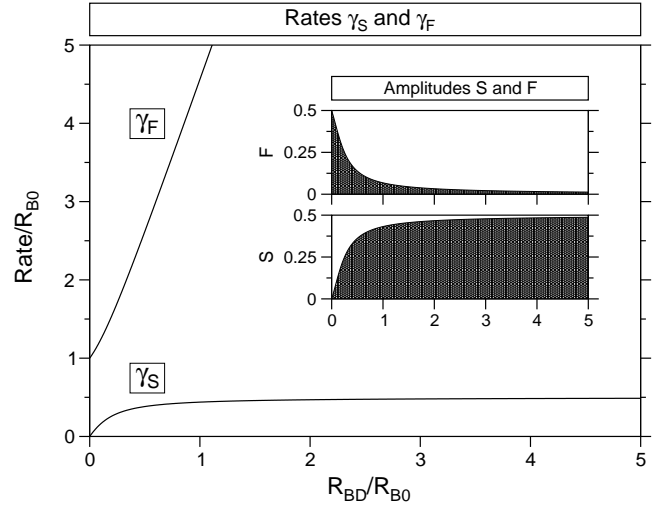}
\caption{{\label{Fig_3}}Fast [$\gamma_F$; Eq. (\ref{gamma_A.solution.Table_1})]
  and slow [$\gamma_S$; Eq.  (\ref{gamma_B.solution.Table_1})] components of
  the monoexciton population $n_{B}$ in units of $R_B$. Inset: Amplitudes $F$
  [Eq. (\ref{amp_A.solution.Table_1})] and $S$ [Eq.
  (\ref{amp_B.solution.Table_1})] versus $R_{BD}$. The rates $R_{B0}$ and
  $R_{D0}$ that enter $\gamma_F$ and $\gamma_S$ are those of Table
  \ref{Table_1}.}
\end{figure}

\noindent In time-resolved PL experiments the measured 
signal $I(t)$ is proportional to the number of photons per unit of time:
$dn_0/dt$ [Eq. (\ref{Eq_3})], which under the assumption of $R_{D0}=0$ results
in $I(t)=R_{B0}\,n_B(t)$. Note that by being proportional to the occupation of
the bright state the signal $I(t)$ carries information on both radiative and
non-radiative (spin flip) processes.
Figure \ref{Fig_2} shows the logarithm of $n_B(t)$ as a function of time for
different spin-flip rates $R_{BD}$, and Fig. \ref{Fig_3} shows the slow
($\gamma_{S}$) and fast ($\gamma_{F}$) components of the decay of $n_B(t)$
versus $R_{BD}$.
We find that in the limiting cases of (1) extremely {\em slow} ($R_{BD}\ll
R_{B0}$) and (2) extremely {\em fast} ($R_{BD}\gg R_{B0}$) spin flip the decay
of $n_{B}$ is primarily determined by a single exponential. 
%
% slow thermalization
%
In case (1) we find $\gamma_F \simeq R_{B0}+2R_{BD}$ and $\gamma_S \simeq
2R_{BD}$, while $F\simeq 1/2$ and $S\simeq 0$ [Fig. \ref{Fig_3}, inset]. Here,
$\gamma_F$ resembles the expected result for the decay rate of the PL in the
presence of nonradiative recombination centers; the dark state of the
monoexciton in this case.\cite{pankove_book,gurioli_PRB_2006}
The population of the bright state is

\begin{equation}
n_B(t)\simeq\frac{1}{2}\exp[-(R_{B0}+2R_{BD})\,t].
\end{equation}

\noindent In this regime, $I(t)$ decays approximately with the characteristic 
lifetime of the bright states.
%
% Fast thermalization
%
In case (2), we find $\gamma_F\simeq\ R_{B0}/2+4R_{BD}$ and $\gamma_S\simeq
R_{B0}/2$, and $S\sim 1/2$ and $F\sim 0$ [Fig. \ref{Fig_3}, inset]; therefore,

\begin{equation}
n_B(t)\simeq\frac{1}{2}\exp\left(-\frac{R_{B0}}{2}\,t\right) .
\end{equation}

\noindent $I(t)$ thus decays with an approximate characteristic time of 
$\tau_R(X^0)=2R_{B0}^{-1}$; {\em twice} as large as the characteristic
lifetime of the bright state.

%%%%%%%%%%%%%%%%%%%%%%%%%%%%%%%%%
%                               %
%   Comparison with experiment  %
%                               %
%%%%%%%%%%%%%%%%%%%%%%%%%%%%%%%%%
%
%
%
\section{Comparison with existing experimental data}

In the experimental literature, data on the bright-dark transition time are scarce. 
{\em (In,Ga)As/GaAs dots.} Dalgarno and
co-workers\cite{dalgarno_pssa_2005,smith_PRL_2005} have studied recently the
effect of the dark state in the decay of the monoexciton in a gated structure
and have estimated $R_{BD}^{-1} > 20 \;{\rm ns}$ at a temperature of $5\;{\rm
  K}$. In addition, they have found that this spin-flip rate varies strongly
with the applied bias. From time-resolved PL experiments, Favero {\em et
  al.}\cite{favero_PRB_2005} have extracted significantly disparate values for
two different dots: $R_{BD}\sim 440$ and $30\;{\rm ns}$.

%
% InP/InGaP
%
{\em InP/(In,Ga)P dots.} In a two-photon absorption experiment, Snoke and
co-workers\cite{snoke_PRB_2004} populated the dark states of the monoexciton
and measured the subsequent luminescence as a function of time. From the data
below $70\;{\rm K}$, the authors found that the spin-flip, bright-dark
transition time satisfies $R^{-1}_{BD}\ge 200\;{\rm ps}$.

%  
% CdSe
%
{\em CdSe dots.} By performing fluorescence transient experiments at room
temperature, Wang {\em et al.}\cite{wang_JPCB_2006} have concluded that the
bright-dark, spin-flip relaxation times $R^{-1}_{BD}=0.2$-$0.4\;{\rm ps}$. Thus,
the spin-flip process is orders of magnitude faster than the characteristic
radiative recombination time.
These experiments reveal order-of-magnitude variations in the bright-dark
spin-flip times, suggesting that the value of $R_{BD}$ appropriate to
interprete time-resolved photoluminescence experiments in quantum dots is
controversial and further research is needed to understand the spin-flip
mechanism.

In the slow spin flip regime, consistent with the findings of Dalgarno {\em et
  al.}\cite{dalgarno_pssa_2005} and Favero {\em et al.}\cite{favero_PRB_2005},
our calculated value $\tau_{R}(X^0)\simeq R^{-1}_{B0}=1.1\;{\rm ns}$ for the
In$_{\rm 0.6}$Ga$_{\rm 0.4}$As/GaAs dot is in excellent agreement with the
data of Bardot {\em et al.}\cite{bardot_PRB_2005}, who extracted $1.55\;{\rm
  ns}$ from time-resoved photoluminescence, and the value of $1\;{\rm ns}$
found by Buckle {\em et al.}\cite{buckle_JAP_1999} and Stevenson {\em et
  al.}\cite{stevenson_PhysicaE_2004}
In the fast spin flip regime, for our prototypical CdSe dot (see Table
\ref{Table_1}), we obtain $\tau_R(X^0)=17\;{\rm ns}$, which is in excellent
agreement with the values of $17\;{\rm ns}$ and $19\;{\rm ns}$ extracted,
respectively, by Brokmann {\em et al.}\cite{brokmann_PRL_2004} and Labeau {\em
  et al.}\cite{labeau_PRL_2003} from time-resolved photoluminescence in
ZnS-passivated CdSe dots.
In the regime of intermediate spin flip rates, measurements of the
biexponential decay of $I(t)$---as those performed by Dalgarno {\em et
  al.}\cite{dalgarno_pssa_2005}---could be used to deduce the spin-flip rate.

%%%%%%%%%%%%%%%%%%%%%%%%%%%%%
%                           %
%   Solution of t(X)/t(XX)  %
%                           %
%%%%%%%%%%%%%%%%%%%%%%%%%%%%%
%
%
%
\section{the ratio $\tau_R(X^0)/\tau_R(XX^0)$}

To compare the biexciton decay rate with to the monoexciton decay rate in the
limiting cases discussed above (Sec. \ref{integration_rate_eq}), we first note
that atomistic pseudopotential-based calculations show (Table \ref{Table_1})
that the characteristic radiative rates for the biexciton ground state satisfy

\begin{eqnarray}
R_{0B}(XX^0) \simeq R_{B0}(X^0) , \nonumber \\ 
\label{relationships_epm}
R_{0D}(XX^0) \simeq R_{D0}(X^0) .  
\end{eqnarray}

\noindent Second, in contrast to $X^0$, we note that the
solution of the rate equation for $XX^0$ [Eq. (\ref{rate_biexciton})]
results in $n_B(XX^0)\sim \exp(-\gamma\,t)$; a single exponential that decays with rate

\begin{equation}
\gamma=2(R_{0B}+R_{0D})\simeq 2(R_{B0}+R_{D0})
\end{equation}

\noindent {\em regardless} of the value of the spin-flip rate 
$R_{BD}$. Similarly to $X^0$, the time-resolved PL signal is proportional to
the population of the bright state of the biexciton.
For slow spin flip [case (1), Sec.
\ref{integration_rate_eq}] we find a decay-rate ratio between $X^0$ and $XX^0$ of

\begin{equation}
\label{X0_slow}
\gamma/\gamma_F \simeq \frac{2R_{0B}}{R_{B0}+2R_{BD}} \simeq 2 \quad\text{(slow spin flip)} ,
\end{equation}

\noindent and for fast spin flip [case (2), Sec. \ref{integration_rate_eq}] we find 

\begin{equation}
\label{X0_fast}
\gamma/\gamma_S\simeq \frac{2R_{0B}}{R_{B0}/2}\simeq 4 \quad\text{(fast spin flip)}.
\end{equation}

We emphasize that depending on the magnitude of the spin-flip time the ratio
$\tau_R(X^0)/\tau_R(XX^0)$ can change by a factor of two, therefore the
assumed spin-flip time is crucial when comparing results for
$\tau_R(X^0)/\tau_R(XX^0)$.
Recently, Wimmer {\em et al.}\cite{wimmer_PRB_2006} have used a quantum
Monte Carlo (QMC) approach with model single-band effective-mass electron and
hole states to calculate 

\begin{equation}
\tau_{R}(X^0)/\tau_R(XX^0)\simeq 2 \quad\text{({\em slow} spin flip)} .
\end{equation}

\noindent Those authors speculated that the disagreement with the
pseudopotential and CI calculations of Ref. \onlinecite{narvaez_PRB_2005c},
which adopt the fast spin-flip regime and predict

\begin{equation}
\label{ratio_fast}
\tau_{R}(X^0)/\tau_{R}(XX^0) \simeq 4 \quad\text{({\em fast} spin flip),}
\end{equation}

\noindent originates from an inaccurate treatment of
correlations in CI. However, as is obvious from Eqs. (\ref{X0_slow}) and
(\ref{X0_fast}), the discrepancy can be directly attributed to the different
assumptions for the spin flip rates. 
%
%Remove before resubmission
%
%Moreover, the ratio $\tau_R(X^0)/\tau_R(XX^0)$ is a good measure of the effect
%of correlation and our agreement with the QMC results of Wimmer {\em et
%  al.}\cite{wimmer_PRB_2006}---when the same spin flip rates are assumed---is
%a confirmation of our accurate treatment of correlations for the purpose of
%contrasting.
%
When the same spin-flip rates are assumed, our results for
$\tau_R(X^0)/\tau_R(XX^0)$ are in agreement with the QMC results of Wimmer
{\em et al.}\cite{wimmer_PRB_2006}, suggesting that our treatment of
correlations is adequate to predict the ratio of monoexciton and biexciton
radiative lifetimes.
On the other hand, the calculated values of the radiative recombination times
depend not only on the treatment of correlation effects, but also on the
accuracy of the electron and hole wavefunctions.
Our atomistic results are in good agreement with experiment while the results
of Wimmer {\em et al.}\cite{wimmer_PRB_2006} based on the single-band effective
mass approximation differ from experimental data by a factor of two.

Finally, note that in our calculation of $\tau_R(X^0)/\tau_R(XX^0)$ we assume
that the the change in occupation of the monoexciton bright and dark states
[Eq.  (\ref{eq:5levels})] is not affected by the decay of the biexciton state.

%%%%%%%%%%%%%%%
%             %
%   Summary   %
%             %
%%%%%%%%%%%%%%%
%
%
%
\section{Summary}

We calculated the characteristic radiative recombination rates for the ground
state of the monoexciton and biexciton in self-assembled (In,Ga)As/GaAs and
colloidal CdSe quantum dots using atomistic wave functions.
For the monoexciton we used these rates in a model three-level system
of rate equations where we varied the spin-flip rate $R_{BD}$.  The latter
affects significantly the radiative decay time: Fast spin flip leads to an
exciton radiative recombination rate twice as fast as the rate obtained from
slow spin flip.
The radiative decay times $\tau_{R}(X^0)$ calculated in the limit of slow spin
flip are in excellent agreement with available data for self-assembled dots,
while for colloidal dots the agreement is best for fast spin flip.
The biexciton radiative decay is a single exponential with a relaxation time
that is independent of the spin-flip rate. But the ratio between the radiative
decay time of the biexciton $\tau_R(XX^0)$ and monoexciton does depend on
$R_{BD}$ and results, respectively, in $\tau_R{X^0}/\tau_{R}(XX^0)\simeq 4$
and $2$ for fast and slow spin flip.
This result resolved the apparent contradiction between the calculation of
Wimmer {\em et al.}\cite{wimmer_PRB_2006}, who predicted
$\tau_R{X^0}/\tau_{R}(XX^0)\simeq 2$ and our previous atomistic
calculation\cite{narvaez_PRB_2005c} in which we found
$\tau_{R}(X^0)/\tau_R(XX^0) \simeq 4$.

\begin{acknowledgments}

This work was funded by the U.S. Department of Energy, Office of Science,
Basic Energy Sciences, under contract No. DE-AC36-99GO10337 to NREL.

\end{acknowledgments}

%%%%%%%%%%%%%%%
%             %
%   APPENDIX  %
%             %
%%%%%%%%%%%%%%%
%
%
%

\appendix

%%%%%%%%%%%%%%%%%%%%%%%%
%  Analytic solution   %
%%%%%%%%%%%%%%%%%%%%%%%%
%
\section{Solution of the three-level model rate equations for the radiative
  decay of $X^0$}
\label{analytic_solution}

We use the simplified model of Eq. (\ref{Eq_3}) as shown in Fig.
\ref{Fig_1}(b). To find $n_B(t)$ and $n_D(t)$, we propose

\begin{equation}
n_B(t)=F\exp(-\gamma_F\,t)+S\exp(-\gamma_S\,t), 
\end{equation}

\noindent with $n_B(0)=F+S$ and $\gamma_F\neq\gamma_S$, and solve first for
$n_D(t)$ with initial condition $n_D(0)$, where $n_B(0)+n_D(0)=1$.

\begin{widetext}

  Then, to solve for $S$, $F$, $\gamma_S$, and $\gamma_F$, we substitute the
  solution of $n_D(t)$ in the rate equation for
  $dn_{B}/dt$ [Eq. (\ref{Eq_3})], obtaining the following conditions.

\begin{eqnarray}
\gamma_F^2-\left(R_{B0}+2R_{BD}+R_{D0}+2R_{DB}\right)\,\gamma_F+
R_{B0} R_{D0} + 2R_{DB} R_{B0} + 2R_{BD} R_{D0}=0 , \nonumber \\
\\
\gamma_S^2-\left(R_{B0}+2R_{BD}+R_{D0}+2R_{DB}\right)\,\gamma_S+
R_{B0} R_{D0} + 2R_{DB} R_{B0} + 2R_{BD} R_{D0}=0 , \nonumber \\
\\
2R_{DB}\,n_{D}(0)+(R_{D0}+2R_{DB})\,n_B(0)-(R_{B0}
R_{D0}+2R_{DB}R_{B0}+2R_{BD}R_{D0})\left(\frac{F}{\gamma_F}+\frac{S}{\gamma_S}\right)=0
.\nonumber \\
\end{eqnarray}

\noindent Thus, we find

\begin{eqnarray}
\label{gamma_A.solution}
\gamma_{F}&=&\frac{1}{2}\left(R_{B0}+R_{D0}+2(R_{BD}+R_{DB})\right)+ \nonumber\\ 
& &
+\frac{1}{2}\sqrt{(R_{B0}-R_{D0})^2+4(R_{B0}-R_{D0})(R_{BD}-R_{DB})+4(R_{BD}+R_{DB})^2}
\\
\label{gamma_B.solution}
\gamma_{S}&=&\frac{1}{2}\left(R_{B0}+R_{D0}+2(R_{BD}+R_{DB})\right)-\nonumber
\\
& & -\frac{1}{2}\sqrt{(R_{B0}-R_{D0})^2+4(R_{B0}-R_{D0})(R_{BD}-R_{DB})+4(R_{BD}+R_{DB})^2}
\end{eqnarray}

\noindent and

\begin{eqnarray}
F&=&\frac{R_{B0}+2R_{BD}-\gamma_S}{\gamma_F-\gamma_S}\,n_B(0)-\frac{2R_{DB}}{\gamma_F-\gamma_S}n_D(0)
\\
S&=&-\frac{R_{B0}+2R_{BD}-\gamma_F}{\gamma_F-\gamma_S}\,n_B(0)+\frac{2R_{DB}}{\gamma_F-\gamma_S}n_D(0)
\end{eqnarray} 
\end{widetext}

\end{document}